\documentclass[fleqn,twoside,twocolumn,nofootinbib]{revtex4} % Specifies the document class %,unsortedaddress
\newcommand{\varkappa}{\xi}
\begin{document}
\title[MODIFIED THERMODYNAMICS AS AN APPROACH TO THE DESCRIPTION]%колонтитул
{MODIFIED THERMODYNAMICS AS AN APPROACH\\ TO THE DESCRIPTION OF
SOME UNIVERSAL\\ PROPERTIES OF ``NEARLY PERFECT FLUIDS''}%
\author{A.D. Sukhanov}%1 автор
%\affiliation{Bogoliubov Laboratory of Theoretical Physics,
%JINR}%институт
%\address{Dubna,
%Russia}%адрес
%\email{ogol@mail.ru}%e-mail
\author{V.G. Bar'yakhtar}%
%\affiliation{National Academy of Science}%
%\address{Kiev,
%Ukraine}%
%\email{vbar@imag.kiev.ua}
\author{O.N. Golubjeva}%
%\affiliation{Peoples' Friendship University of Russia}%
%\address{Moscow, Russia}%
%\email{ogol@oldi.ru} \udk{530.1} \pacs{67.10.Fj;  47.10.-g}
%\razd{\secix}

\setcounter{page}{1309}%

\begin{abstract}
We show that the quantum statistical mechanics describing quantum
and thermal properties of objects has only the sense of a particular
semiclassical approximation. We propose a more general (than that
theory) microdescription of objects in a heat bath taking a vacuum into
account as an object environment; we call it
$(\hbar,k)-$dynamics ($\hbar kD)$. We introduce a new generative
operator, a Schr\"{o}dingerian or a stochastic action operator, and will show
its fundamental role in the determination of such macroquantities as
internal energy, effective temperature, and effective entropy. We
establish that $\hbar kD$ can serve as an initial microtheory for
constructing a modified thermodynamics. On this ground, we can
explain the universality of the ratio ``effective action to effective
entropy'' at zero temperature and its minimal value in the form
$\hbar/2k$. This result corresponds to experimental data obtained
recently under studying a new matter state -- a nearly perfect
fluid.
\end{abstract}

\maketitle

\section{Experimental Data and the Status of the Theory of Nearly Perfect Fluids}

In the latest years, a new direction of experimental researches
has started to be developed. It has a concern with the joint analysis of the
entropy density and the shear viscosity behavior in a wide temperature
interval in various media: in Bose-liquids of the helium-4 type,
in ultracold Bose- and Fermi-gases in traps, in traces of
quark-gluon plasma at collisions of heavy nuclei, and even in
graphene [1--5].

It is proved that, in so different media, the universal behavior of the
ratio of the shear viscosity $ \eta $ to the volume density of entropy $s$
at $T\rightarrow 0$ or $T\rightarrow T_{\rm cr}$ is observed. We
note that the limiting value of the ratio is proportional to the
ratio of the Planck and Boltzmann world constants. In addition, the obvious
dependence of this ratio on the temperature is observed. We assume
that it is not a casual result, but some evidence on the existence of
a qualitatively new matter state under conditions of the strong
interaction. The properties do not depend on specific features
of the initial  medium, but they are completely defined by
the quantum-thermal influence of an environment. In the literature, this
state has received the name ``nearly perfect fluid'' [1].

As one can see from the table below, the minimal value of the ratio
$\eta$/${s}$ in units of the ratio $\hbar$/$k_{\rm B}$ is near to 0.5
for various media [5]:\vskip3mm
    \tabcolsep22.2pt
\noindent{\footnotesize\begin{tabular}{l|l}

          \hline
         \multicolumn{2}{c}{Viscosity/entropy ratio: current status }\\
        \hline
           Quark-gluon plasma   &   $0<\eta/s< 0.5$  \\
         %\hline
           Trapped cold alkali atoms  &  $(\eta/s)_{\min}\approx 0.5$   \\
         %\hline
           Liquid helium  &   $(\eta/s)_{\min}\approx 0.7$  \\
         \hline

      \end{tabular}}
\vskip3mm

The idea of joint consideration of the quantum and thermal stochastic
influences of an environment was implicitly present at Planck's and
Einstein's  papers at the  beginning of the XX-th century. However, till the 1970s,
its realization in the well-known theories of Neumann, Matzubara, {\it etc.}
was connected with the separate and independent account of the both
types of stochastic influences. The feature of these theoretical
results consists in the fact that the minimal value of entropy is
equal to zero, and fluctuations of the temperature are absent. But many
experiments abundantly evidence the another.

We assume that the reason for this discrepancy lies in the inconsistency
of the theories. The matter is that they
use a classical model of the heat bath as a set of weakly
connected classical oscillators in especially quantum problems. Therefore, those theories
omit the essentially quantum influence of a vacuum in
the presence of the oscillator zero-point energy.

At a later time, N.N.~Bogolyubov has suggested [6] (1978) to replace a
classical model of the heat bath by the quantum one. This idea has
opened a possibility to consider simultaneously the quantum and
thermal types of stochastic influence, but it has not been used by his
followers in a wide class of problems. Practically simultaneously, a
close idea of thermofield vacuum has stated by Umezawa [7] who offered to
describe it by a wave function with regard for the presence of a thermal
noise. At the same time in the thermofield dynamics developed by
him with employees, the definition of entropy
corresponding to quantum statistical mechanics (QSM) was still used, and
temperature fluctuations were not considered.

However, by the end of the XX-th century, some additional data have been obtained in a number of experiments
which testify to an essential role
of temperature fluctuations and a nonzero minimal value of entropy.
These facts initiated searches of the theories giving the best
consent with experiment.

In particular, in work [1], a theoretical scheme has been first
offered, in which the joint account of quantum-thermal influences
allowed one to get a nonzero value of entropy and a
temperature-independent limiting value of the ratio $ \eta/s$. On
its basis, a model of the medium with ``nearly perfect fluidity''
as a certain  collective substance was constructed. As a
qualitatively essential element, the given theoretical scheme uses
the relativistic quantum field theory at finite temperatures and
the general relativistic theory. This model has been actively
developed in publications of other authors on the basis of other
approaches such as the kinetic theory, numerical modeling, and the
theory of holographic duality [2,~3].

In the theory [1] developed for the model of quark-gluon plasma with
strong interaction, a number of results  does not correspond to
experimental data for various  media: first, the ratio
$\eta/s$  does not depend on the temperature; second, the minimal
value of the ratio equals 0.08, which strongly differs from
the experimental result of 0.5 quoted in the same work.

Thus, now there is a certain progress in the explanation of a limiting
value of the ratio $ \eta/s$ for quark-gluon plasma. However, the
high-grade theory suitable for an adequate explanation of the whole set
of observable facts, including the universality of the given
limiting ratio, is absent. Considering that these phenomena concern
with the area of quantum-thermal ones, we suggest to address to the
thermodynamic approach as the most universal sight at the nature.
However, for our goals, the thermodynamics itself needs some
modification.

The conviction that the traditional thermodynamics is completely
derived from the equilibrium quantum statistical mechanics (QSM) has
predominated for a quite long time. It is assumed that QSM allows
establishing the observable interrelations between macroparameters (
thermodynamics laws, equations of state, {\it etc.}) [8].  But it is
well known that there exist such macroparameters (the temperature, for
example), whose analogs have not yet been studied on the
microlevel.

Nowadays, QSM is not able to
calculate quantum and thermal phenomena simultaneously and
completely. As is well known, QSM is based on the notion of the
density matrix (operator). In the energy representation, it has
the form of the Gibbs--von Neumann quantum canonical distribution,
%1
\begin{equation}
w_n=\exp\frac{F-\varepsilon_n}{\Theta},
\end{equation}
where $\varepsilon_n$ is the spectrum of the object energy, $F$
is the free energy determined by the normalization condition, and
$\Theta^{-1}$ is the modulus of the distribution.

We note the next limitations:

\noindent 1. Insensitivity of Gibbs' distribution to the oscillator
zero-point energy $\varepsilon_0=$ $\hbar\omega/2$:
%2
\begin{equation}\label{1a}
\left.\begin{array}{lcl}
\varepsilon_n\Rightarrow\varepsilon_n^\prime=\varepsilon_n+\varepsilon_0
\\
F_n\Rightarrow F_n^\prime=F_n+\varepsilon_0\\
\end{array}\right\}
F^\prime-\varepsilon_n^\prime=F-\varepsilon_n.
\end{equation}

\noindent 2. Moreover, as can be seen in the ground state of the
quantum oscillator, we have
%3
\begin{equation}\label{2}
\Delta p_0 \Delta q_0 =\frac{\hbar}{2}
=\frac{\varepsilon_0}{\omega},
\end{equation}
which confirms the direct relation between the quantities
$\varepsilon_0 $ and $\hbar$/2. (It is principal in quantum
physics!)

\noindent 3. In QSM, it is assumed that the object temperature does
not fluctuate. The Zero law is:
%4
\begin{equation}
T=T_0;\;\;\Delta T=0,
\end{equation}
while the temperature fluctuations in low-temperature experiments
are sufficiently noticeable for small objects, including
nanoparticles and the relict radiation.

 \noindent 4. In addition, in QSM,
Gibbs' distribution yields automatically the Third law:
%5
\begin{equation}
 S_{\min}=0.
 \end{equation}
By referring to experiments, we may assert that the zeroth minimum entropy is
currently very doubtful.

According to Gibbs' distribution, the expression
%6
\begin{equation}
\Theta_{\rm cl} = k_{\rm B} T= \varepsilon_{\rm cl}
\end{equation}
corresponds in QSM to choosing the classical model of the heat bath as a
set of weakly coupled classical oscillators  (Krylov, Bogoliubov
(1939) [9]). But a microobject with quantized energy can be placed
in such a heat bath even under the conditions $k_{\rm
B}T<(\varepsilon_n-\varepsilon_{n-1})$. At the same time, the idea of
the additivity of quantum and thermal contributions contradicts
the thermal radiation theory.

\section{ A Programm of Constructing of Modern Stochastic
Thermodynamics}

 So, as the equilibrium QSM is not a consistent theory for
constructing the equilibrium thermodynamics suitable for describing
the low-temperature area and small objects, we proceed from another
microtheory that takes quantum-thermal effects into account, in
particular, temperature fluctuations. It is the so-called $(\hbar
k)$-dynamics [10]. Instead of the apparatus of density matrix, it is
grounded on the complex wave function, whose amplitude and phase
depend on the temperature. The suggested  theory  does not have the
defects inherent in QSM.

On the base of  this theory, we introduced a new macroparameter --
the effective temperature, in which  the constants $\hbar$ and
$k_{\rm B}$ play equipollent roles. This temperature allows one to
more fully describe the thermal equilibrium state, but it presumes
using another model of heat bath. So, we start from the next
statement:\looseness=1

Our goal is constructing  a new  macrotheory -- the modern
stochastic thermodynamics based on the evident account of the stochastic
influence that is characterized on the microlevel with the
stochastic action operator.

We hope that the modified thermodynamics will allow us to more closely
approach the answers to the following unresolved questions:\\
  Why does the minimal value of the ratio $ \eta$/$s$
appear universal for the state of a nearly perfect fluid in so various
media?\\
 Whether such universal ratio follows from the fundamental
principles of the known standard theories -- quantum mechanics and the
statistical mechanics?\\
 Whether it is necessary to use specific
features of quantum field theory and general relativity theory
in the description of a nearly perfect fluid in extremely various
media?

\section{ A Model of Quantum  Heat bath. The ``Cold'' Vacuum}

In constructing the new  microtheory ($\hbar kD$) as the base of
the modified thermodynamics, we  consider that no objects are
isolated in the nature. In this connection, we follow the \emph{Feynman
approach} [11], according to which any system can be represented as
a set of the object under study and its environment (the ``rest of
the Universe''). The environment can exert both regular and
stochastic influences on the object. Here, we study only the
stochastic influences. Two types of influences, namely, quantum and
thermal actions characterized by the respective Planck and Boltzmann
constants can be assigned to it.

So, to obtain a consistent quantum-thermal description of natural
objects, we use the approach, in which we can modify the fundamental
microdescription of the objects under thermal equilibrium
conditions. For these purposes, we propose to formulate  \emph{a
quantum-thermal dynamics} or, briefly,
a ($\hbar k$)-dynamics, as a modification of the standard quantum mechanics taking
thermal effects into account. The principal distinction of such a
theory from QSM is that the state of a microobject in it under the
conditions of contact with a quantum-heat environment is generally described
not by the density matrix, but by \emph{a temperature-dependent
complex wave function.}

We note that this is not a ``technical sleight-of-hand''. Using  the
wave function, we thereby suppose to consider pure and mixed states
simultaneously in the frame of \emph{Gibbs' ensemble}. It is, in
principle, differs from \emph{ Boltzmann' assembly} used in QSM.

To construct our theory, it is necessary:

1. \emph{To change}   $ \hat\rho(T) \Rightarrow \Psi_T(q) $.

2. To introduce (except the Hamiltonian) a new operator -- the
\emph{stochastic action operator}  $\widehat{\mathfrak S}$.

3. To use an idea of heat bath at $T=0$ (\emph{``cold''  heat
bath}).

4. To use an idea of vacuum at $T>0$ (\emph{``thermal''} vacuum)
also.

This theory is based on a new microparameter, namely, the stochastic
action operator.  In this case, we demonstrate that, by averaging the
corresponding microparameters over the temperature-dependent wave
function, we can find the \emph{most important effective
macroparameters}, including the internal energy, temperature, and
entropy. They have the physical meaning of the standard
thermodynamic quantities typical of a phenomenological
macrodescription.

To describe the environment with the holistic stochastic influences,
we introduce a specific model, the quantum heat bath (QHB) [12].
According to this, the QHB is a \emph{set of weakly coupled quantum
oscillators} with all possible frequencies. The equilibrium thermal
radiation can serve as a preimage of such a model in the nature.

The specific feature of our understanding of this model is that
we must apply it to both the ``thermal'' ($T\neq  0$) and the
``cold'' ($T = 0$) vacua. Thus, in the sense of Einstein [13] and
Bogoliubov [6], we proceed from a more general understanding of the
thermal equilibrium which can, in principle, be established for any
type of environmental stochastic action (\emph{purely quantum,
quantum-thermal, and purely thermal}).

We begin our description  by studying the ``cold'' vacuum and
discussing the description of a single quantum oscillator from the
number of oscillators forming the QHB model for $T= 0$ from a new
standpoint.

But we recall that the lowest state in the energetic
($\Psi_n(q)$) and coherent states (CS) is the same. In the $q$
representation, the same ground state of a quantum oscillator is,
in turn, described by the \emph{real wave function}
%7
\begin{equation}
\Psi_0(q) =[2\pi (\Delta q_0)^2]^{-1/4}\exp \left\{-\frac{q^2}
{4(\Delta q_0)^2}\right\}.
\end{equation}
In the occupation number representation, the ``cold'' vacuum in
which the number of particles is $n=0$ corresponds to this state.

As is well known, coherent states (CS) are the eigenstates of the
non-Hermitian particle annihilation operator $\hat a$ with complex
eigenvalues. But they include one isolated state $|0_a\rangle =
|\Psi_0(q)\rangle$ of the particle vacuum, in which eigenvalue of
$\hat a$ is zero
%8
\begin{equation}
\hat{a}|0_a\rangle=0|0_a\rangle;\quad \hat{a}|\Psi_0(q)\rangle
=0|\Psi_0(q)\rangle.
\end{equation}

In what follows, it is convenient to describe the QHB in the $q,$ but
not $n$ representation. Therefore, we express the annihilation
operator $\hat a$ and the creation operator $\hat a^\dagger$  in
terms of the operators $\hat p$ and $\hat q,$ using the
traditional method. We have
%9
\[\hat a=\frac1 2 \left (\frac{\hat p}{\sqrt{\Delta
p_0^2}}-i\frac{\hat q}{\sqrt{\Delta q_0^2}}\right);
\]
\begin{equation}
\hat a^\dag=\frac1 2 \left (\frac{\hat p}{\sqrt{\Delta
p_0^2}}+i\frac{\hat q}{\sqrt{\Delta q_0^2}}\right),
\end{equation}
where $(\Delta q_0)^2 =$ $\hbar$/2$m\omega$ and
$(\Delta p_0)^2 =\hbar m\omega$/2\\

The particle number operator then becomes
%10
\begin{equation}
 \hat N_a=\hat a^\dag\hat a=\frac{1}{\hbar\omega}\left(\frac{\hat
p^2}{2m}-\frac{\hbar\omega}{2}\hat I+ \frac{m\omega^2\hat
q^2}{2}\right).
\end{equation}
After multiplying this relation by
$\hbar\omega$, we obtain the standard interrelation between the
expressions for the Hamiltonian in the $q$ and $n$ representations:
%11
\begin{equation}
\hat{\mathcal{H}}=\frac{\hat p\,^2}{2m}+\frac{m\omega^2\hat
q^2}{2}=\hbar\omega(\hat N_a+\frac12\hat I),
\end{equation}
where $\hat I$  is the unit operator.

From the thermodynamics standpoint, we are concerned with the
internal energy of the quantum oscillator in equilibrium with the
``cold'' QHB. Its value is equal to
%12
\[ U_0=\langle\Psi_0(q)|\hat{\mathcal H}|\Psi_0(q)\rangle=
\]
\begin{equation}
 =\hbar\omega
\langle\Psi_0(q)|\hat
{N}_a|\Psi_0(q)\rangle+\frac{\hbar\omega}{2}=\frac{\hbar\omega}{2}
=\varepsilon_0.
\end{equation}

In the given case, the state without particles coincides with the
state of the Hamiltonian with the minimum energy $\varepsilon_0$.
\emph{The quantity} $\varepsilon_0$ traditionally treated as the
oscillator zero-point energy takes the physical meaning \emph{of
the internal energy }$U_0$  of a quantum oscillator in equilibrium
with the ``cold'' vacuum.

\section{A Model of Quantum  Heat Bath. The ``Thermal'' Vacuum}

We can pass from the ``cold''  to the ``thermal''  vacuum using the
Bogoliubov $(u, v)$-transformation with the
\emph{temperature-dependent coefficients} [12]
%13
\[u=\left(\frac12\coth\frac{\hbar\omega}{2k_{\rm B}T}+\frac12\right)^{1/2}e^{\textstyle
{i\frac{\pi}{4}}};
\]
\begin{equation}
v=\left(\frac12\coth\frac{\hbar\omega}{2k_{\rm
B}T}-\frac12\right)^{1/2}e^{-\textstyle {i\frac{\pi}{4}}}.
\end{equation}

In the given case, this transformation is canonical but leads to a
unitary nonequivalent representation, because the QHB at any
temperature is a system with an infinitely large number of degrees of freedom.
Such a transformation allows passing from the set of CS to
a more general set of states called the \emph{thermal
correlated coherent states}. They are selected because they ensure
that the Schr\"{o}dinger coordinate - momentum uncertainties relation
for a quantum oscillator is saturated at any temperature.

From the second-quantization apparatus standpoint,  the
transformation ensures the passage from the original system of
\emph{particles}  with the ``cold'' vacuum  $|0_a\rangle$ to the
system of \emph{ quasiparticles} described by the annihilation
operator $\hat b$ and the creation operator $\hat b^{\dag}$ with the
``thermal'' vacuum $|0_b\rangle$.

So, we pass

1. From cold  to thermal  vacuum:
\[\Psi_0(q)\Rightarrow\Psi_T(q); \quad
|0_a\rangle\Rightarrow|0_b\rangle.\]

2. From particles to quasiparticles:
\[\hat a \Rightarrow \hat b =
\hat b(T); \quad\hat a^\dag \Rightarrow \hat b^\dag = \hat
b^\dag(T).\]

The choice of transformation coefficients is fixed by the
requirement that the expression for the mean energy of a quantum
oscillator in the thermal equilibrium be defined by the Planck formula,
which can be obtained from experiments:
%14
\begin{equation}
 \varepsilon_{\rm Pl}=\hbar\omega(\exp{\frac{\hbar\omega}{k_{\rm B}T}}-1)^{-1}
+\frac{\hbar\omega}{2}=\frac{\hbar\omega}{2}\coth\frac
{\hbar\omega}{2k_{\rm B}T}=\varepsilon_{qu}.
\end{equation}

Earlier, it was shown by us [12] that the state of the \emph{``thermal''
 vacuum} $|0_b\rangle
\equiv|\Psi_{\scriptscriptstyle{T}}(q)\rangle$ in the $q$
representation corresponds to the complex wave function
%15
\begin{equation}
\Psi_{\scriptscriptstyle{T}}(q)=[2\pi (\Delta q)^2]^{-1/4}\exp
\left\{-\frac{q^2} {4(\Delta q)^2}(1-i\alpha)\right\},
\end{equation}
where
%16
\begin{equation}
 (\Delta q)^2 = \frac{\hbar}{2m\omega} \coth\frac
{\hbar\omega}{2k_{\rm B}T};\quad\alpha
=\left[\sinh\frac{\hbar\omega}{2k_{\rm B}T}\right]^{-1}.
\end{equation}
For its Fourier transform $\Psi_{\scriptscriptstyle{T}}(p),$ a
similar expression with the same coefficient $\alpha$ and $(\Delta
p)^2 = \frac{\hbar m\omega}{2} \coth\frac {\hbar\omega}{2k_{\rm
B}T}$ holds.

\section{Some Features of  the ``Thermal'' Vacuum }

Of course, the states from the set of thermal correlated coherent
states are the eigenstates of the non-Hermitian quasiparticle
annihilation operator  $\hat b$ with complex eigenvalues. They also
include one isolated state of the \emph{quasiparticle vacuum}, in
which eigenvalue of $\hat b$ is zero,
%17
\begin{equation}
 \hat{b}|0_b\rangle=0|0_b\rangle; \;\;\;\;\hat{b}|\Psi_T(q)\rangle
=0|\Psi_T(q)\rangle.
\end{equation}
Using the wave function of the ``thermal'' vacuum, we obtain the
expression for the operator $\hat b$ in the $q$ representation:
%18
\[\hat b=\frac {1}{2}\sqrt{\coth
\frac{\hbar\omega}{2k_{\rm B}T})}\times
\]
\begin{equation}
 \times\left[\frac{\hat
p}{\sqrt{\Delta p_0^2}}-i\frac{\hat q}{\sqrt{\Delta q_0^2}} (\coth
\frac{\hbar\omega}{2k_{\rm B}T})^{-1}(1-i\alpha)\right].
\end{equation}
The corresponding quasiparticle creation operator $\hat b^{\dag}$
can be obtained from $\hat b$ with $i$ replaced by $(-i)$ \\
%19
\[ \hat b^{\dag}=\frac {1}{2} \sqrt{\coth
\frac{\hbar\omega}{2k_{\rm B}T})}\times
\]
\begin{equation}
\times\left[\frac{\hat p}{\sqrt{\Delta p_0^2}}+i\frac{\hat
q}{\sqrt{\Delta q_0^2}} (\coth \frac{\hbar\omega}{2k_{\rm
B}T})^{-1}(1+i\alpha)\right].
\end{equation}
We can verify that, as $T\rightarrow 0,$ the operators $\hat
b^{\dag}$ and $\hat b$ for quasiparticles pass to the operators
$a^{\dag}$ and $\hat a$ for particles, and
\[|0_b\rangle\Rightarrow
|0_a\rangle;\;\;\;\;\Psi_T(q)\Rightarrow\Psi_0(q).\]

Acting just as above, we obtain the expression for the
\emph{quasiparticle number operator} $\hat N_b$ in the $q$
representation
%20
\[ \hat N_b=\hat b^{\dag}\hat b=\frac 14\coth
\frac{\hbar\omega}{2k_{\rm B}T}\times
\]
\begin{equation}
\times\left[\frac {\hat p^2}{\Delta p_0^2}-2(\coth
\frac{\hbar\omega}{2k_{\rm B}T})^{-1} (\hat I+\frac \alpha
\hbar\{\hat p,\hat q\})+\frac {\hat q^2}{\Delta q_0^2}\right],
\end{equation}
where we take $1+\alpha^2=\coth^2\frac{\hbar\omega}{2k_{\rm B}T}$
into account when calculating the last term.

Multiplying by $\hbar\omega$ and passing from the quasiparticle
number operator to the original Hamiltonian, we obtain
%21
\begin{equation}
 \hat{\mathcal{H}}=\hbar\omega(\coth
\frac{\hbar\omega}{2k_{\rm B}T})^{-1}\left[\hat N_b+\frac 12 \hat I+
\frac\alpha \hbar\{\hat p,\hat q\}\right].
\end{equation}
We stress that the operator $\{\hat p,\hat q\}$ in this formula can
also be expressed in terms of bilinear combinations of the operators
$\hat b^{\dag}$ and $\hat b$, but they differ from the quasiparticle
number operator. This means that the operators $\hat{\mathcal H}$
and $\hat N_b$ do \emph{not commute}, and the wave function
$\psi _T(q)$ is therefore \emph{not the eigenfunction of the
Hamiltonian} $\hat{\mathcal{H}}$.

As before, we are interested in the thermodynamic quantity, namely,
the internal energy $U$ of a quantum oscillator now in thermal
equilibrium with the ``thermal''  QHB. Calculating it just as
earlier, we obtain
%22
\[U= \hbar\omega(\coth
\frac{\hbar\omega}{2k_{\rm B}T})^{-1}\times
\]
\begin{equation}
\times\left[\langle\Psi_{\scriptstyle T}(q)|\hat
N_b|\Psi_{T}(q)\rangle + \frac{ 1}{2 }+\frac{\alpha}{2\hbar}
\langle\Psi_{\scriptstyle T}(q)|\{\hat p,\hat q\}
|\Psi_{T}(q)\rangle\right]
\end{equation}
in the $q$ representation.

Because we average over the quasiparticle vacuum in this formula,
the first term in it vanishes. At the same time, it was shown
earlier by us [14] that
%23
\begin{equation}
\langle\Psi_{\scriptstyle T}(q)|\{\hat p,\hat
q\}|\Psi_{T}(q)\rangle=\hbar\alpha.
\end{equation}
As a result, we obtain the expression for the internal energy of the
quantum oscillator in the ``thermal'' QHB in the $\hbar kD$:
%24
\begin{equation}
 U=\frac {\hbar\omega}{2 (\coth
\frac{\hbar\omega}{2k_{\rm B}T})}(1+\alpha^2)
=\frac{\hbar\omega}{2}\coth \frac{\hbar\omega}{2k_{\rm
B}T}=\varepsilon_{\rm Pl}.
\end{equation}

This means that the average energy of the quantum oscillator at
$T\neq 0$ has the thermodynamic meaning of its internal energy in
the case of equilibrium with the ``thermal'' vacuum. As
$T\rightarrow 0$, it passes to a similar quantity corresponding to
the equilibrium with the ``cold'' $\;$ vacuum.

\section{The Stochastic Action Operator -- Schr\"{o}dingerian}

Because the original statement of the $\hbar kD$ is the \emph{idea
of the holistic stochastic action} of the QHB on the object, we want
to introduce a new operator in the Hilbert space of microobject
states to implement it.

In this connection, we recall the Schwartz inequality
%25
\begin{equation}
|A|^2\cdot|B|^2\geq|A\cdot B|^2,
\end{equation}
the Schr\"{o}dinger uncertainty relation (SUR) for the coordinate and the momentum
following from it [15]:

1. Unsaturated SUR
%26
\begin{equation}
(\Delta p)^2(\Delta q)^2 > |\tilde
R_{qp}|^2\equiv\sigma^2+\frac{\hbar^2}{4}.
\end{equation}

2. Saturated SUR
%27
\begin{equation}
 (\Delta p)^2(\Delta q)^2 = |\tilde R_{qp}|^2.
 \end{equation}
In the absent of a stochastic action, $\tilde{R}_{qp}\equiv0.$ For real
states, $\sigma = 0$.

Hereinafter, we use an analysis of the right-hand side of the
saturated coordinate-momentum SUR. For not only a quantum
oscillator in a heat bath but also for any object, the complex
quantity on the right-hand side of the SUR,
%28
\begin{equation} \widetilde{R}_{p\,q}=\langle\Delta p|\Delta
q\rangle\quad\mbox{or  \quad                      }
\widetilde{R}_{p\,q}=\langle\,|\Delta \widehat{p}\,\Delta
\widehat{q}\,|\,\rangle,
\end{equation}
has a double meaning.

On the one hand, it is the \emph{amplitude of the transition} from the
state $|\Delta q\rangle $ to the state $|\Delta p\rangle $. On the
other hand, it can be treated as the \emph{  quantum correlator}
calculated over an arbitrary state $|\; \rangle$  of some
operator. We note that it has a dimension of action.

The \emph{nonzero value} of the quantity $\widetilde{R}_{pq}$ is
the \emph{fundamental attribute of nonclassical theory}.
Therefore, it is quite natural to assume that the averaged
operator in that formula has a fundamental meaning. In view of
dimensional considerations, we call it \textbf{the stochastic
action operator or Schr\"{o}dingerian},
%29
\begin{equation}
 \widehat{\mathfrak S}\equiv\Delta \widehat p\,
\Delta \widehat{q}.
\end{equation}
At the first time, it was
introduced in the paper by E. Schr\"{o}dinger (1930) [16]. Of course, it
should be remembered that the operators $\Delta \hat q $ and
$\Delta\hat p $ do not commute, and their product is a non-Hermitian
operator.

We can express  the given operator in the form
%30
\begin{equation}
 \widehat{\mathfrak S}= \frac{1}{2} \left\{
\Delta\widehat{p}\Delta\widehat{q}+
\Delta\widehat{q}\Delta\widehat{p}\,\right\}
+\frac{1}{2}\left[\Delta\widehat{p}\Delta\widehat{q}-
\Delta\widehat{q}\Delta\widehat{p}\,\right]=
 \widehat{\sigma}-i\,\widehat{\mathfrak S}_0.
 \end{equation}
It allows separating the Hermitian part of $\widehat{\mathfrak S}$
from the anti-Hermitian one. Then the Hermitian operators
$\widehat\sigma$ and $\widehat{\mathfrak S}_0$ have the form
%31
\begin{equation}
 \widehat{\sigma}\equiv \frac12\{\Delta \hat p,\;\Delta \hat q\};
\quad\widehat{\mathfrak S}_0 \equiv \frac{ i}{2}\;
[\hat{p},\hat{q}]= \frac{\hbar}{2}\;\hat {I}.
\end{equation}
It is easy to see that the mean
$\sigma=\langle\,|\widehat{\sigma}|\,\rangle$ of the operator
$\hat\sigma$ resembles the expression for the standard correlator of
coordinate and momentum fluctuations in classical probability
theory. It transforms into this expression if the operators $\Delta
\hat q $ and $\Delta \hat p$ are replaced with $c$-numbers. It
reflects the contribution of the environmental stochastic action to the transition amplitude $\tilde
R_{pq}$.

Therefore, we call the operator $\hat{\sigma}$ \emph{the external
action operator} in what follows. Previously, the possibility to
use a similar operator $\hat\sigma$ was discussed by Krylov and
Bogoliubov [9], where it was studied as a \emph{quantum analog} of
the  action variable in the set of action-angle  classical
variables.

At the same time, the operators $\widehat{\mathfrak S}_0$ and
$\widehat{\mathfrak S}  $ were \emph{not previously introduced}. The
operator $\widehat{\mathfrak S}_0$ reflects a specific peculiarity
of the objects to be ``sensitive'' to the minimum stochastic action
of the ``cold'' vacuum and to respond to it adequately regardless of
their states.

Therefore, it should be treated as a \emph{minimal stochastic action
operator}. Its mean
%32
\begin{equation}
 \langle \;|\hat{\mathfrak S}_0|\,\rangle\equiv J_0=
\frac {\hbar}{2}
\end{equation}
is independent of the choice of the
state, over which the averaging is performed, and it has, hence, the
meaning of the invariant eigenvalue of the operator
$\widehat{\mathfrak S}_0$.

This implies that, in the given case, we deal with the
universal quantity $J_0$ which is called the minimal action. Its
fundamental character is already defined by its relation to the
Planck world constant $\hbar$. But the problem is not settled yet.

Indeed, according to the tradition dating back to Planck, the
quantity $\hbar$ is assumed to be called the elementary quantum of
action. At the same time, the factor 1/2 in the quantity $J_0$
plays a significant role, while half the quantum of the action is
not observed in the nature.

Therefore, the quantities $\hbar$ and $ \frac 12 \hbar $, whose
dimensions coincide, have different physical meanings and must
be named differently, in our opinion. From this standpoint,
it would be more natural to call the quantity $\hbar$ the external
action quantum. Hence, the quantity $\hbar$ is the minimum portion of the action
transferred to the object from the environment or from another
object. Therefore, photons and other quanta of fields being carriers
of fundamental interactions are first the carriers of the minimal
action equal to $\hbar$. The same is also certainly related to
phonons.

Finally, we note that only the quantity $\hbar$ is related to the
discreteness of the spectrum of the quantum oscillator energy in the
absence of the heat bath. At the same time, the quantity $ \frac
\hbar 2$ has an independent physical meaning. It specifies  the
minimum value of the macroparameter -- the internal energy $U_0$ of
the quantum oscillator in the ``cold'' QHB (at $T=0$).

Below, we wil evaluate the specific features of the Schr\"{o}dingerian used in the
microdescription. We recall that this operator is
non-Hermitian. This fact would seemingly contradict the standard
requirements imposed on the operators in quantum mechanics, but
there is nothing unusual in this.

If we are interested in genuine quantum
dynamics which is naturally associated with transitions from one
state to another one, then precisely the non-Hermitian
operators play an important role. The creation and annihilation
operators or, for example, the scattering matrix are among the most
well-known of them. The stochastic action operator $\widehat
{\mathfrak S}$ also belongs to these operators.

\section{Effective Action as a Fundamental Macroparameter}

We now construct the macrodescription of objects using their
microdescription in the $\hbar kD$. The mean $\widetilde {\mathfrak
S}$ of the operator $\widehat {\mathfrak S }$  coincides with the
complex transition amplitude $\widetilde R_{pq}$ or quantum
correlator and, in thermal equilibrium, can be expressed as
%33
\begin{equation}
 \widetilde {\mathfrak S}=\langle \Psi_{\scriptstyle
T}(q)|\,\widehat{\mathfrak S}\,|\Psi_{\scriptstyle
T}(q)\rangle=\sigma - \,i\,J_0,
\end{equation}
where $\sigma $ and $J_0$  are the means of the corresponding
operators. In what follows, we regard the modulus of the complex
quantity  $\widetilde {\mathfrak S}$,
%34
\begin{equation}
  |\widetilde{\mathfrak S}
|=\sqrt{\sigma^2+J_0^2}=\sqrt{\sigma^2+\frac{\hbar^2}{4}}\equiv
J_{\rm ef}
\end{equation}
as a new macroparameter -- the effective action. For the quantum
oscillator, it has the form
%35
\begin{equation}
 J_{\rm ef}=\frac\hbar 2\coth\frac{\hbar\omega}{2k_{\rm B}T}
 \end{equation}
 and coincides with a quantity
previously postulated from intuitive considerations [12].

Now we establish the interrelation between the effective action and
traditional thermodynamic quantities. Comparing the expressions for
$J_{\rm ef}$ and $U$, we can see that $U=\omega|\tilde{\mathfrak
S}|=\omega J_{\rm ef}$. In the high-temperature limit, where $
\sigma\rightarrow \sigma_T= k_{\rm B}T/\omega\gg \hbar/2,$ this
relation becomes $U=\omega \sigma_T$. This formula was obtained
by Boltzmann [17] (1904) for macroparameters in classical
thermodynamics by generalizing the concept of adiabatic
invariants  in classical mechanics.

The formula obtained above also allows expressing the interrelation
between the effective action $J_{\rm ef}$ and the effective
temperature $T_{\rm ef}$ [18]  in the explicit form:
%36
\begin{equation}
T_{\rm ef}=\frac{\varepsilon_{qu}}{k_{\rm B}}=\frac{\hbar\omega}
{2k_{\rm B}}\coth\frac{\hbar\omega}{2k_{\rm
B}T}=\frac{\omega}{k_{\rm B}}J_{\rm ef}.
\end{equation}
This yields
%37
\begin{equation}
 T^{min.}_{\rm ef}= \frac{\omega}{k_{\rm B}}\mathfrak S_0=
\frac{\hbar\omega}{2k_{\rm B}}\neq 0.
\end{equation}

\section{Effective Entropy in the \boldmath$\hbar kD$}

The possibility of introducing the entropy in the $\hbar kD$ is also
based on using the wave function instead of the density operator.

Using the corresponding dimensionless expressions
$\tilde{\rho} (\tilde q)$ and $\tilde{\rho}(\tilde p)$ instead of $\rho (q)= |\Psi_T(q)|^2$ and $\rho(p) =
|\Psi_T(p)|^2$, we
propose to define a formal coordinate -- momentum entropy $S_{qp}$
by the equality
%38
\begin{equation}
 S_{qp}\,=-k_{\rm B}\left\{\int\tilde{\rho}(\tilde{q})
\ln\tilde{\rho}(\tilde{q})
d\tilde{q}+\int\tilde{\rho}(\tilde{p})\ln\tilde{\rho}(\tilde{p})
d\tilde{p}\right\}.
\end{equation}

Substituting the corresponding dimensionless expressions and using
the normalizing condition, we obtain
%39
\begin{equation}
S_{qp}= k_{\rm B} \left\{(1+\ln\frac{2\pi}{\delta})
+\ln\coth\frac{\hbar\omega}{2k_{\rm B}T}\right\}.
\end{equation}
The final result depends on the choice of the constant $\delta$. If
$\delta = 2\pi$, we can interpret the given expression as the
quantum-thermal entropy or, briefly,  $S_{QT},$ because it coincides
exactly with
%40
\begin{equation}
S_{QT} \equiv S_{\rm ef} =k_{\rm
B}\left\{1+\ln\coth\frac{\hbar\omega}{2k_{\rm
B}T}\right\}\rightarrow k_{\rm B},
\end{equation}
obtained earlier by us in the macrotheory
framework [12].

This expression, as distinct from the entropy in QSM, takes
the averaging both of the amplitude $\overline{|\Psi_T
(q)|}$ and the phase $\overline{\varphi}=\frac \alpha 4$ into account.

\section{Standard Thermodynamics on the Base of the Effective
Action}

Using the $\hbar kD$, we can introduce the effective action $J_{\rm
ef} $ as a new fundamental macroparameter. Its advantage
is that it has a \emph{microscopic preimage} --  the
stochastic action operator $\widehat{\mathfrak S}$ or
Schr\"{o}dingerian. Moreover, in thermal equilibrium, we can express
the main thermodynamic characteristics of objects -- temperature and
entropy -- in terms of it.

For example, we  recall that $T_{\rm ef}\sim J_{\rm ef}$. It follows
that $J_{\rm ef}$ is also an intensive macroparameter
characterizing the stochastic action of the ``thermal'' QHB.

In view of this, the \emph{Zero law} of equilibrium modern
stochastic  thermodynamics instead of
%41
\begin{equation}
 T_{\rm ef}= T_{\rm ef}^{\rm therm} \pm \Delta T_{\rm ef},
\end{equation}
can be rewritten as
%42
\begin{equation}
 J_{\rm ef}= J_{\rm ef}^{\rm therm} \pm
\Delta J_{\rm ef},
\end{equation}
 where $J^{\rm therm}_{\rm ef}$ is effective action of QHB, and $J_{\rm ef}$ and $\Delta
J_{\rm ef}$ are, respectively, the means of the effective action of an object and
the standard deviation from it. The state of thermal equilibrium can
actually be described, in such cases, in the sense of Newton,
assuming that
\begin{center}
\emph{``the stochastic action is equal to the stochastic
counteraction''.}
\end{center}

We now turn to the effective entropy. In the absence of a mechanical
contact,
%43
\begin{equation}
 dS_{\rm ef}= \frac{\delta Q_{\rm ef}}{T_{\rm ef}}=
 \frac{dU}{T_{\rm ef}}.
 \end{equation}
Now we can rewrite this expression in the form
%44
\begin{equation}
 dS_{\rm ef}=k_{\rm B}d(\ln\frac{J_{\rm ef}}{\mathfrak S_0})
\equiv dS_{QT}.
\end{equation}

It follows that the effective or $QT$-entropy, being an
extensive macroparameter, can be also expressed in terms of $J_{\rm
ef}$. Correspondingly,  \emph{ the third law} is in accordance with
Nernst initial statement
%45
\begin{equation}
 S_{QT}= S_{\rm ef}=
k_{\rm B}\{1+\ln\coth\frac{J_{\rm ef}}{J_0}\}\rightarrow  k_{\rm
B}\ne 0.
\end{equation}

\section{Summary}

Nowadays, there is a number of experiments and theories where the
ratio $\eta/s$ is the subject of interest. It was shown by us [12]
that this quantity can be given by
%46
\begin{equation}
 \frac{\eta}{s}\equiv\frac{J^{\rm ef}}{S^{\rm ef}}= \varkappa\;
 \frac{\coth\varkappa\omega/T}{1+\ln\coth
\varkappa\omega/T}\rightarrow \varkappa,
\end{equation}
where
%47
\begin{equation}
 \varkappa\equiv\frac{J^{\rm ef}_{\min}}{S^{\rm ef}_{\min}}
=\frac{\hbar}{2 k_{\rm B}}=3.8\times 10^{-12}\;\; \mbox{K$\cdot$ c}
\end{equation}
and $J_{\min}^{\rm ef}= J_0\equiv$ $ \hbar/2$,  $S^{\rm ef}_{\min}=
k_{\rm B}$ are the limiting values for $T\ll T^{\rm ef} $.

In contrast to the $\hbar kD$, QSM yields
%48
\begin{equation}
 \frac{\eta}{s}\rightarrow\frac{\hbar\exp(-\hbar\omega/k_{\rm B}T)}
{k_{\rm B}(\hbar\omega/k_{\rm B}T)\exp(-\hbar\omega/k_{\rm
B}T)}=\frac{T}{\omega}\rightarrow 0.
\end{equation}
So, it is possible to compare two theories ($\hbar kD$ and QSM)
experimentally by measuring the limiting value of this ratio: the
ratio $\frac{\eta}{s}$ is equal to either $\varkappa$ or zero.

So, we have proposed the ($\hbar k$)-dynamics as a more general
microdescription of objects in a heat bath with the vacuum
explicitly taken into account than that presented by QSM. We have constructed a
basically new model of the object environment, namely, a quantum
heat bath. We have studied its properties including the cases of
``cold'' and ``thermal'' vacua.

We have introduced a new stochastic action operator -- Schr\"{o}dingerian --
and shown its fundamental role in the microdescription. We
have established that the corresponding macroparameter, the effective
action $J_{\rm ef}$, plays just a significant  role in the
macrodescription. The most important effective macroparameters of
equilibrium quantum statistical thermodynamics such as the internal energy,
temperature, and entropy are expressed in terms of this
macroparameter.

So, we have found, in part, the answer to the questions: Why is this
ratio universal for different nearly perfect fluids? Does this ratio
follow from the general principles of quantum mechanics and
statistical mecha\-nics?

\vskip3mm
 We suppose that, yes. The ratio is an universal one, and it
follows from the general principles of not quantum mechanics or
statistical mechanics taken separately, but of a more generalized
joint theory -- the $(\hbar, k)$-dyna\-mics.\looseness=1

As a result, it appears that the stochastization process of object
characteristics due to the contact with QHB is described at the macrolevel
by two qualitatively various quantities -- effective entropy and
effective temperature. Nevertheless, at any absolute temperatures,
they can be expressed through  the same macroparameter -- effective
influence. It is important to notice that, in the $(\hbar, k)
$-dynamics, the last value has a microscopic prototype. It is the
stochastic influence operator or Schr\"{o}dingerian depending on
the world constants of Planck and Boltzmann.

 So, it is necessary to stress that the theory offered above, the $ (\hbar, k)
$-dynamics, not only has already allowed explaining the
universality of the limiting value of the ratio $ \frac {\eta} {s}$
observed in experiments. It has established a microscopic prototype
of such a quantity as the temperature. As we have pointed earlier,
QSM is not able to do it, because there is not a suitable operator
like the Schr\"{o}dingerian in it.  At its further generalization to
nonequilibrium wave functions, the given theory can also play the
important role in the explanation of various properties of nearly
perfect fluids.

However, the area of its applicability can be wider. The
fundamental idea of the transition from a classical model of the heat
bath to a quantum model automatically leads to the necessity of
the replacement
\[ k_{\rm B} T\rightarrow k_{\rm B} T _ {ef} \]
in the majority of the formulas deduced  earlier within the limits
of thermodynamics and statistical mechanics.\looseness=1

Doing this replacement, we open a possibility of a deeper analysis
of the existing experiments under conditions of equilibrium with QHB
at ultralow temperatures. In this case, some quantum effects can be
revealed, by obviously reflecting the presence of quantum stochastic
influences of the ``cold''  vacuum. Certainly, the discovery of such
effects will allow interpreting, in a new manner, many already known
quantum effects shown at ultralow temperatures.\looseness=1

\vskip3mm This work was supported by the Russian Foundation for
Basic Research (Project No.~10-01-90408 ).

\end{document}